# Fairness and Stability Analysis of Congestion Control Schemes in Vehicular Ad-hoc Networks

Neda Nasiriani, Yaser P. Fallah
Lane Department of Computer Science and Electrical
Engineering, West Virginia University, USA
Email: nnasiria@mix.wvu.edu; yfallah@csee.wvu.edu

Hariharan Krishnan
Integrated Electrical and Control Lab
General Motors global R&D, Warren, MI, USA
Email: hariharan.krishnan@gm.com

*Abstract*— Cooperative vehicle safety (CVS) systems operate based on broadcast of vehicle position and safety information to neighboring cars. The communication medium of CVS is a vehicular ad-hoc network. One of the main challenges in large scale deployment of CVS systems is the issue of scalability. To address the scalability problem, several congestion control methods have been proposed and are currently under field study. These algorithms adapt transmission rate and power based on network measures such as channel busy ratio. We examine two such algorithms and study their dynamic behavior in time and space to evaluate stability (in time) and fairness (in space) properties of these algorithms. We present stability conditions and evaluate stability and fairness of the algorithms through simulation experiments. Results show that there is a trade-off between fast convergence, temporal stability and spatial fairness. The proper ranges of parameters for achieving stability are presented for the discussed algorithms. Stability is verified for all typical road density cases. Fairness is shown to be naturally achieved for some algorithms, while under the same conditions other algorithms may suffer from unfairness issues. A method for resolving unfairness is introduced and evaluated through simulations.

## I. Introduction

One of the most important applications of vehicular networks is vehicle-to-vehicle communications for Cooperative Vehicle Safety (CVS) [1]. In these systems vehicles frequently broadcast their position and safety information to allow hazard prediction. This information is transmitted over a shared wireless channel; receiving vehicles use this information to keep track of their neighboring vehicles and be aware of unsafe situations ("Fig. 1"). The CVS system relies on Dedicated Short Range Communications (DSRC) and WAVE as its communication and networking medium [2][3][4]. The DSRC channel dedicated to CVS is a 10 MHz channel. The safety application may access this channel at any time if a dual-radio solution is used, otherwise the safety channel is available only for a portion of the time on a periodic basis (referred to as channel switching in IEEE 1609 standards) [4]. The initial recommendations in [5] for broadcast of safety messages suggested sending messages with rate of 10pkt/sec and to distance of 150m to 250m.

The initial design was shown to result in channel congestion and scalability issues when a large number of vehicles are in range of each other [11][8]. With CVS being considered for large scale deployment, there has been a lot of efforts in resolving the scalability issue through congestion management schemes [8][6][9][11]. Several algorithms are under field study at the time of this writing, including a variant of [8].

The scalability solutions mainly targeted adaptive methods of setting the rate or range of transmission, instead of using the fixed values of 10Hz and 250m [5]. For example in [12] authors presented a way to allocate transmission power to all nodes in a highway in a way that maximizes the minimum amount of transmission power of every node, assuming a maximum target load. Another approach was proposed in [7] to use model based estimators at sender side to reduce the frequency of message dissemination while preserving tracking accuracy of the system. In [8] a joint rate-power control mechanism (using concepts from [7]) has been introduced and compared to the de-facto solution of [5]. [6] presented an alternative design for the power control component of [8].

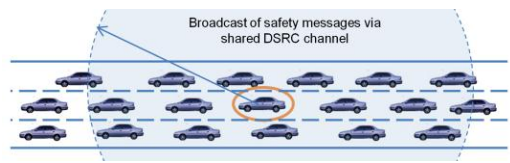

Fig. 1. V2V CVS Communication

While the performance of our proposed algorithms in [8] and [6] have been studied using tracking accuracy and network performance metrics, a thorough analysis of their dynamic behavior (time stability and space fairness) have not been reported. Such an analysis is required, given that a variant of [8] is currently under field study. In this paper we present the mentioned analysis, in particular for the range control (power control) component of the algorithms in [8] and [6].

Algorithms in [8] and [6], as well as method proposed by other researchers in[9], rely on Channel Busy Ratio (CBR) as a network congestion metric, and adapt the broadcast rate or range accordingly. The local value of CBR can be computed by a vehicle using clear channel assessment (CCA) reports from the physical layer to MAC layer in 802.11. Averaging the CCA reading in time, a node can calculate the local CBR. CBR may also be averaged over space, since congestion is a special concept. CBR is considered as a limited feedback measure from the network as it gives clues to how the network is operating, but does not exactly specify its performance without knowledge of other parameters [12].

In this paper we examine the dynamic behavior of the schemes presented in [6][8]. We derive convergence requirements for the range control algorithm proposed in [8]. Such an analysis has been already reported for the method in [6]. In addition to convergence, we discuss the fairness of each algorithm and demonstrate how to adapt range based on local and distributed measurement of Channel Busy Ratio (CBR). We present a method to take advantage of distributed feedbacks from neighboring vehicles to enhance the range control algorithm and achieve better fairness in the network. The next sections provide a background on the congestion management schemes presented in [6][8].

## II. CONGESTION CONTROL ALGORITHMS BASED ON CHANNEL BUSY RATIO

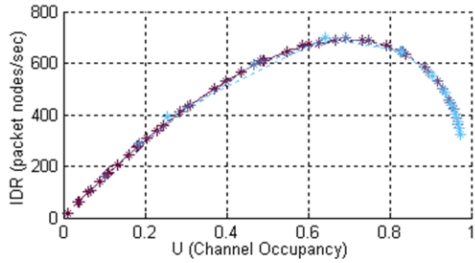

Fig. 2. IDR vs. channel occupancy for different values of r(5-115 msg/sec), d(20-400m), and ρ(0.1-0.2 vehicle/m). Points belonging to the same experiment with different values of d are connected by dotted line and different colors; although due to overlap they are indistinguishable.[6]

We had studied CVS vehicular network in [10] and [12] using a performance measure called information dissemination rate (IDR). IDR is defined as the number of copies of a packet delivered per unit time from a single vehicle to its neighbors up to a given distance $d_{max}$. It was observed that the IDR vs. channel occupancy (CBR) is independent of $r$ (rate), $d$ (range) or even $\rho$ (density) and for all cases it is a dome shaped curve as seen in "Fig. 2".

This shape is dependent on the type of protocol (CSMA/CA in the VANET) and the interference factors (mainly due to hidden nodes in the case of CVS). Assuming that maximizing IDR would benefit the CVS application, and given that rate control was designed based on tracking accuracy requirements, [6] and [8] proposed range control schemes that either try to maximize IDR or maintain it near its maximum. Adaptive algorithms, based on CBR as feedback, have been proposed for range control. These algorithms are described below.

### A. Linear Range Control Algorithm

The idea in this algorithm is to decrease range when congestion is detected, and increase it when network is sensed to be empty. In terms of IDR and CBR, this translates to maintaining CBR or network busy-ness at a level that is near optimal IDR, using a range (power) control scheme that decreases range as CBR increases. "Fig. 3" shows an example of such a function (solid line in red).

We also plot network characteristic curves (describing CBR vs. range) in "Fig. 3" which are plotted based on extensive NS-3 simulation runs for different typical rate and road density values. The control function plotted in "Fig. 3" intersects all the curves in a range of CBR that yields good IDR. When the iterative control algorithm converges, it should stop at the intersection with the current network characteristic curve.

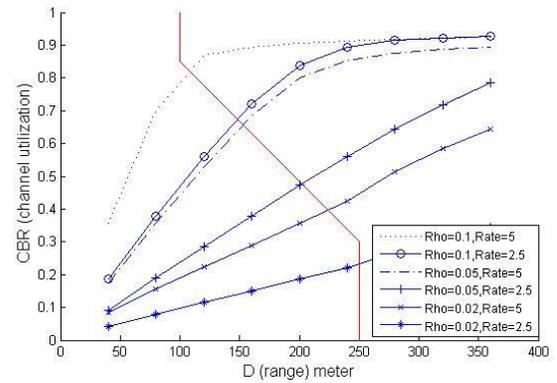

Fig. 3. Communication characteristic curves for six different scenarios, and feedback control function for range control using selected set of limits (4)

There are many choices for controller function formulated as: $D_{k+1} = f(U_k)$ and each will have different intersection with the network characterization graph $U_k = g(D_k)$. The following function is introduced as Linear Range Control (LRC) algorithm based on [8]:

$$D_{k+1} = f(U_k) = \begin{cases} D_{max} & U_k < U_{min} \\ D_{min} + \frac{U_{max} - U_k}{U_{max} - U_{min}}(D_{max} - D_{min}) & U_{min} \leq U_k < U_{max} \\ D_{min} & U_{max} \leq U_k \end{cases} \quad (1)$$

$U_k$ indicates the value of CBR used for adaptation algorithm and the limits for range $(D_{min}, D_{max})$ are obtained from the safety requirements which is desirable to be in range of (50-100) for minimum and (250-300) for the maximum range [10]. In order to define limits for $(U_{min}, U_{max})$, looking at "Fig. 2" we can see that the range should be in range of (0.4-0.8) in order to keep the IDR near its peak value and maintain a good performance in terms of throughput [10].

## B. Gradient Descent Range Control

Looking at "Fig. 2" it can be observed that independent of the network parameters, if the CBR is around $U^* = 0.7$ optimal IDR can be achieved. Therefore, an alternative idea for range control could be to maintain range at a level that results in maximum IDR. This can be achieved through an adaptive range control scheme with an update equation that resembles gradient descent methods. We call this scheme Gradient descent Range Control (GRC) algorithm. GRC tries to maintain CBR at its optimum value defined as $U^*$. The update equation is as follows:

$$D_{k+1} = \min(D_{max}, \max(D_{min}, D_k + \eta(U^* - U_k))) \quad (2)$$

where $\eta$ is the gain that is determined in [6] for a feedback linearized version of the above algorithm; the max and min functions are only used to maintain the range between device and safety limits. Informally, the update equation can be presented as: $D_{k+1} = D_k + \eta(U^* - U_k)$.

## III. STABILITY AND FAIRNESS ANALYSIS

To verify stability of the presented algorithms, we have done a large number of simulation experiments in NS-3. An 8-lane highway is considered, with 4-lane in each direction with identical densities. There are 2000 nodes scattered randomly, 250 in each lane and a set of densities (0.02, 0.05, 0.1) from free flow to very slow traffic are selected for different scenarios. The rate of message transmission was assumed fixed (2.5Hz or 5Hz)in order to allow more intimate study of the range control schemes ; the fixed rate values were taken from typical output rates of the rate control algorithm in [8], which is around 2-3Hz on average and occasionally goes up to 5Hz (on an averaged basis). So we select both 2.5 and 5Hz as typical rate values.

### A. Analysis of the Linear Range Control Scheme

In this section, we study the convergence properties of LRC considering the set of limits introduced in previous section. Here we present lemma 1 which defines the condition for convergence:

Lemma 1: Assuming that network density and average transmission rate stay unchanged, any range control algorithm that uses a decreasing function of CBR ($D_{k+1} = f(U_k)$) is stable in time and converges to a single value for range if the following condition is satisfied:

$$|g(D_{k+1}) - g(D_k)| < |f^{-1}(D_{k+1}) - f^{-1}(D_k)| \quad (3)$$

The convergence property follows from the fact that for the algorithm to converge, two subsequent steps of the algorithm in (1) should lead to smaller difference in the subsequent observed range or CBR. This can be easily visualized in "Fig. 3", and also be written for CBR as

$$|U_{k+1} - U_k| < |U_k - U_{k-1}|. \quad (4)$$

Using equations $D_{k+1} = f(U_k)$ and $U_k = g(D_k)$ and substituting in (4), (3) is derived.

While (3) describes the condition, it needs to be interpreted for different types of controllers $D_{k+1} = f(U_k)$. For LRC, (3) can be simply interpreted as $f^{-1}$ being steeper than $g$ in the entire range of values for $D$ ($D_{min}$ to $D_{max}$).

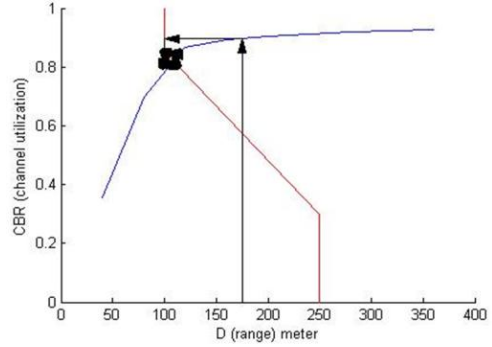

Fig. 4. Convergence study on communication characteristic curve for $\rho = 0.1$ and rate=5 scenario in MATLAB.

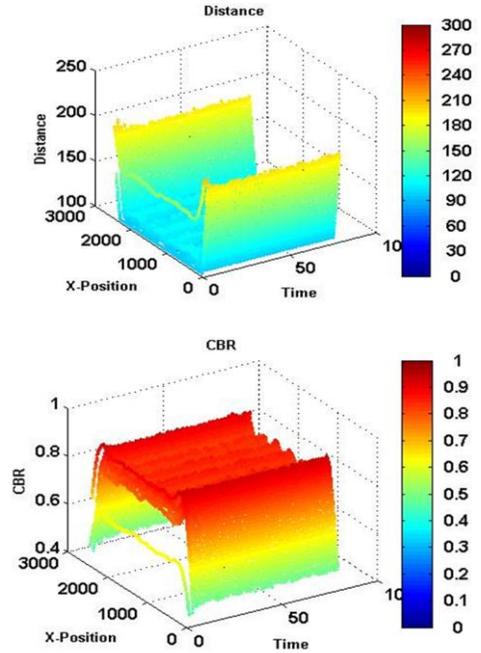

Fig. 5. LRC algorithm for $\rho = 0.1$ and rate=5 scenario with limits as defined in (4) , the algorithm is quick to converge in time and fair in space; top) result showing the distance chosen by each node. bottom) result showing the CBR sensed by each node.

We verified the above condition in a set of simulation runs in MATLAB, using characteristic curves that were derived from NS-3 simulations. With these simulations and having the ideal range and CBR intervals [6][8] we have come up with a set of bounds to define the controller that will converge fairly quick for all scenarios. "Fig. 4" shows the adaptation iterations for $\rho = 0.1$ and rate=5 scenario, with controller parameters set as in (5). This set of parameters for LRC was found to result in convergence in all the considered scenarios. We used these parameters in our NS-3 simulations.

$$\begin{cases}(D_{min},D_{max}) = (100,250) \\ (U_{min},U_{max}) = (0.3,0.85)\end{cases} \quad (5)$$

For network simulations in NS-3, we used an OFDM PHY for the 5 GHz band with 10 MHz channel bandwidth. The size of each packet was set to 500 bytes, sent at 5Hz or 2.5Hz based on the scenario being studied. Nodes use random traffic generation with the specified rate, which describes the case that adaptive control is being used and the sender will decide whether to send a packet or not based on vehicle movement. All the nodes will start randomly and since the broadcast is random too, it is the closest state to the reality.

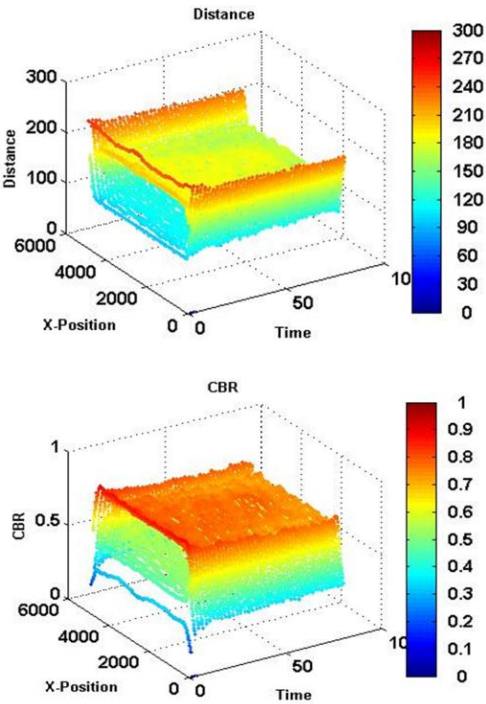

Fig. 6. LRC algorithm for $\rho$ =0.05 and rate=5 scenario with limits as defined in (4) , the algorithm is fairly quick to converge in time and fair in space top) result showing the distance chosen by each node. bottom) result showing the CBR sensed by each node.

With the selected parameters of (5), the LRC algorithm worked perfectly for all the scenarios. As it is seen in "Fig. 5" and "Fig. 6", the algorithm converged quickly over time and remained stable. This is what we were expecting from the MATLAB simulations done in advance. We also simulated a road density change scenario in which the density $\rho$ changes from 0.02 to 0.1 in the middle of the highway (1500m). The algorithm manages to maintain time stability and space fairness as is seen in "Fig. 7".

With the selected parameters, LRC seems to provide a good choice for controlling the range and consequently controlling the congestion over the network. However, we should keep in mind that the selected controller should satisfy (3), which will limit our choices for $D_{min}, D_{max}$ and $U_{min}, U_{max}$. For example by reducing the slope of the control function and changing the limits to (50,300) for range and (0.4,0.8) for CBR, we cannot achieve convergence anymore (see "Fig. 8" and "Fig. 9" for example). If such limits are not acceptable in a specific design, an alternative is to use the Gradient descent Range Control (GRC) algorithm which does not put any limits on maximum or minimum range. Convergence properties of GRC were discussed in [6] and it was recommended that gain ($\eta$) should be selected in a way that 2 conditions are met: 1) range $d$ converges quickly to near optimal value, before the value of $\rho$ or average value of rate changes considerably. 2) the system does not overshoot too much or oscillate and stray into a region that yields significantly low value of IDR (e.g. $u>0.95$ or $u<0.3$) [6]. In this algorithm the range maximum and minimum can be set more liberally and we have chosen 100-300 meter which will be discussed in detail in the next section of the paper.

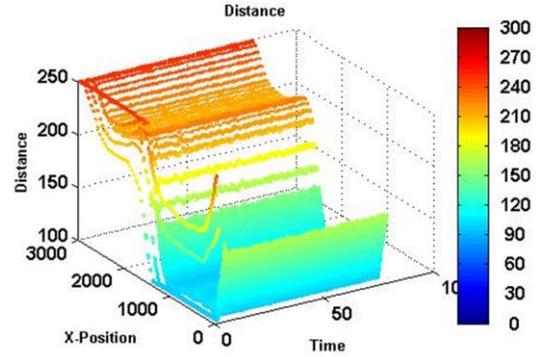

Fig. 7. LRC algorithm result for a mixed scenario which has a $\rho$ =0.1 up to 1500 meter of road and $\rho$ =0.02 for the next half. Stability in time and space can be observed.

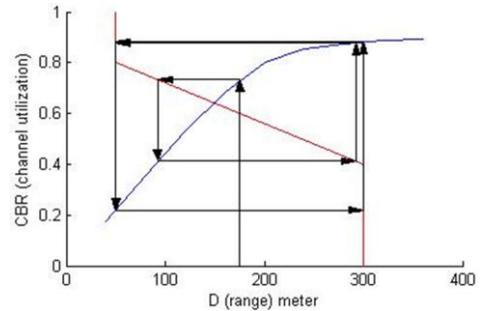

Fig. 8. Convergence Study on communication characteristic curve for $\rho$ =0.05 and rate=5. LRC does not converge in this case.

### B. Gradient Descent Range Control

We first study this algorithm in MATLAB simulation, using network characteristic curves that were obtained from NS-3, to find a good value of $\eta$ that satisfies the conditions explained in previous part ([6] could also be used for a feedback linearized version of GRC). A range of values (10-200) were found to be appropriate.

As expected, using a higher gain $\eta$ (still satisfying the above conditions) will result in faster convergence; however, we observed in NS-3 simulations that although time stability was quickly reached, the system gets quickly into an unfair situation in space. This happens because of the edge effect that propagates inside. By choosing smaller values for $\eta$ convergence happens more slowly (still less than 10 seconds [6]), but system acts more solidly and unfairness in space can now be controlled by a distributed CBR measurement approach. After an extensive set of experiments, we found a gain value around 50 to be appropriate. It can be observed in "Fig. 10" that this value will support convergence to the optimal value ($u^* = 0.7$) in less than 10 seconds. This time is reasonable since change in density and other conditions of the road can seldom happen in shorter time. Based on this observation we have chosen $\eta = 50$ for our GRC simulations in NS-3.

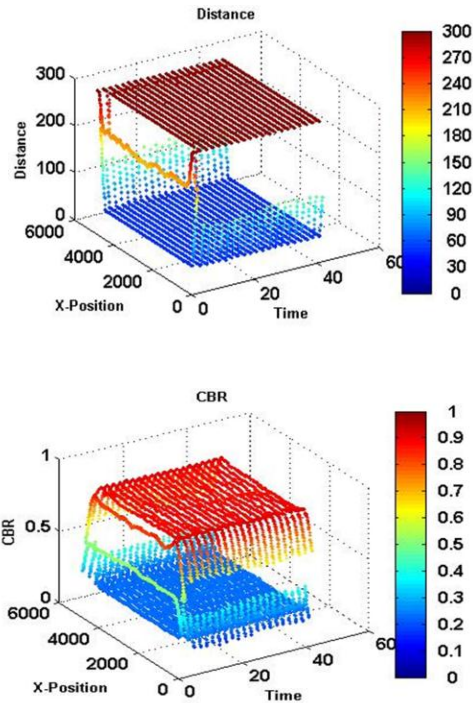

Fig. 9. LRC algorithm for $\rho$ =0.05 and rate=5 scenario with limits $(D_{min}, D_{max})$=(50,300) $(U_{min}, U_{max})$=(0.4,0.8), the algorithm does not converge in time top) result showing the distance chosen by each node. bottom) result showing the CBR sensed by each node.

For very dense and free flow scenarios ($\rho = 0.1, 0.02$) GRC worked perfectly and stability in time and space was observed. However for $\rho = 0.05$ and rate=5, we saw unfairness in space which is shown in "Fig. 11". We believe this is in fact the edge effect that propagates inward from the two edges. Edge nodes will have a lower local CBR sensed and set their range to maximum, causing higher CBR readings for the neighboring nodes. If this happens before other neighboring nodes are operating near their optimal point, they will sense the higher CBR and keep their range to lower values, even to $D_{min}$ in some cases. This phenomenon will propagate inward through the network as seen in "Fig. 11", creating a ripple of low and high range (or high and low CBR), while road density is the same (thus all nodes should have had the same range). In order to solve this issue we tried lower values for $\eta$ to see how the algorithm will behave. Lowering the value of gain (e.g., to $\eta$ =20) would result in this propagation of unfairness to be much slower as seen in "Fig. 12". However, a more suitable solution could be found from distributed measurement of CBR as explained next.

### C. Local vs. Distributed measurment of CBR

In order to solve the issue of unfairness in space, we have introduced a mechanism which every node will consider all measurements of CBR by its neighbors as well as its own locally measured CBR. This distributed method is named "Averaging" in which every node will send the sensed CBR of its channel along with the safety packets to its neighbors (a one byte number). This value is added to the safety message and no extra protocol is required. Using neighboring node's CBR, each node can have a wider picture of the network, which can be helpful in resolving the unfairness issue. We chose to use a simple method of averaging over all heard CBR values from neighbors. The result is depicted in "Fig. 13" and shows that the unfairness issue is resolved.

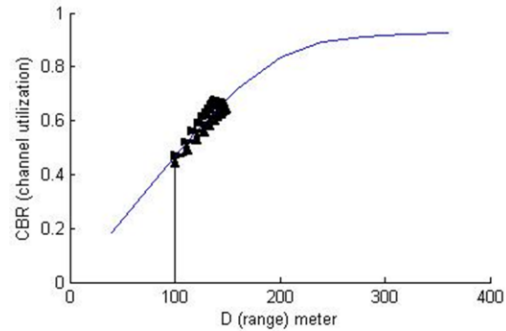

Fig. 10. convergence steps for GRC algorithm for scenario with $\rho = 0.05$ rate=5 and $\eta = 50$ in MATLAB.

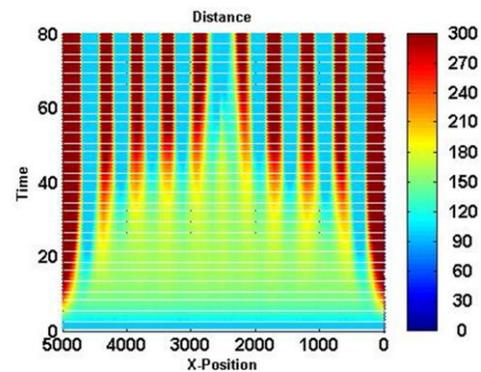

Fig. 11. Result of GRC algorithm with $\eta = 50$ for case $\rho = 0.05$ and rate=5. unfairness is observed in this case.

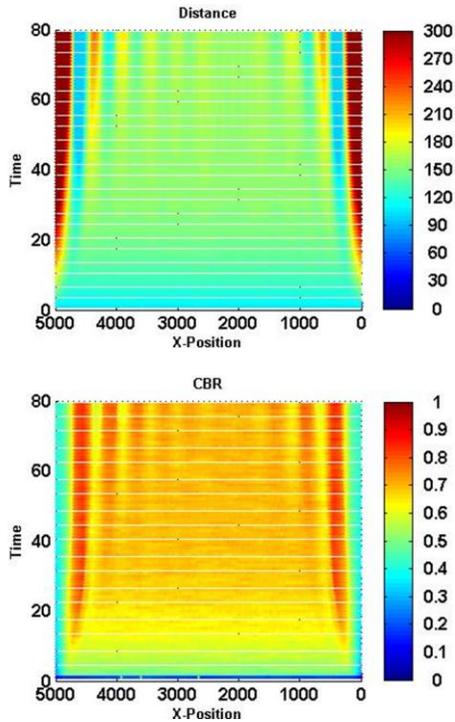

Fig. 12. Result of GRC algorithm with $\eta = 20$ for case $\rho = 0.05$ and rate=5 top) Distance chosen by the algorithm for each node bottom) CBR sensed by each node is shown.

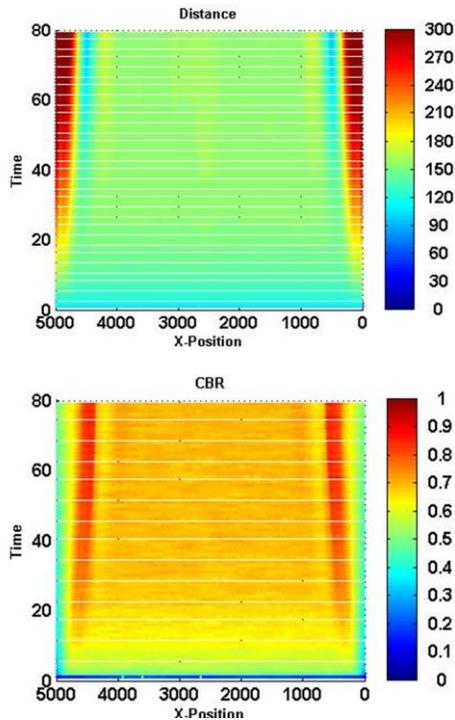

Fig. 13. Result of GRC algorithm with averaging, $\eta = 20$ for case $\rho = 0.05$ and rate=5 top) Distance chosen by the algorithm for each node is shown bottom) CBR sensed by each node is shown.

## IV. CONCLUSION

Cooperative vehicle safety systems are perhaps the most important and challenging application of VANETs at this time. These systems need to be robust enough in case of large scale implementation. Vehicular network size may become extremely large on congested roadways; hence scalability is an important factor in designing such systems. To achieve scalability, congestion management schemes based on adaptive rate and range transmissions have been proposed. In this paper we have examined the dynamic behavior of two of these algorithms to verify their stability and fairness properties. It was found that for the studied algorithms controller function parameters have to meet certain restrictions in order for the algorithm to be stable. In addition, it was also found that fairness issues, which existed for one algorithm, could be alleviated if distributed measurement of network congestion measure (channel busy ratio) was employed.